\pdfoutput=1

\documentclass[11pt,twoside,a4paper,cmspaper,final,collab]{cms-tdr}
\def\svnVersion{83987b3-D}\def\svnDate{2020/08/24}\def\cmsCernNoTag{CERN-EP-2020-149}\def\cmsCernDate{\today}\def\cmsMessage{"Published in the Journal of High Energy Physics as \href{http://dx.doi.org/10.1007/JHEP09(2020)149}{\doi{10.1007/JHEP09(2020)149}.}"}

\begin{document}\cmsNoteHeader{SUS-19-013}

\hyphenation{had-ron-i-za-tion}
\hyphenation{cal-or-i-me-ter}
\hyphenation{de-vices}
\newcommand{\njets}{\ensuremath{N_{\text{jet}}}\xspace}
\newcommand{\mlsp}{\ensuremath{m(\PSGczDo)}\xspace}
\newcommand{\mgluino}{\ensuremath{m(\PSg)}\xspace}
\newcommand{\squark}{{\PSQ}\xspace}
\newcommand{\gjets}{{{\cPgg}\!+jets}\xspace}
\newcommand{\zjets}{{{\cPZ}\!+jets}\xspace}
\newcommand{\dR}{\ensuremath{\Delta R}\xspace}
\newcommand{\dRZb}{\ensuremath{\Delta R_{\PZ,\PQb}}\xspace}
\newcommand{\mjet}{\ensuremath{m_{\operatorname{jet}}}\xspace}
\newcommand{\imini}{\ensuremath{I}\xspace}
\newcommand{\dphimht}{\ensuremath{\Delta\phi_{j,\,\htvecmiss}}\xspace}
\newcommand{\rscale}{\ensuremath{\mu_{\mathrm{R}}}\xspace}
\newcommand{\fscale}{\ensuremath{\mu_{\mathrm{F}}}\xspace}
\newcommand{\invfb}{\fbinv}
\newcommand{\met}{\ptmiss}
\newcommand{\NLSP}{\PSGczDt}
\newcommand{\LSP}{\PSGczDo}
\newcommand{\znn}{\ensuremath{\PZ\to \PGn\PAGn}\xspace}
\newcommand{\lsp}{\LSP}
\newcommand{\nlsp}{\NLSP}
\newcommand{\wjets}{\ensuremath{\PW\text{+jets}}\xspace}
\newcommand{\Bnorm}{\ensuremath{\mathcal{B}_{\operatorname{norm}}}\xspace}
\newlength\cmsFigWidth
\ifthenelse{\boolean{cms@external}}{\setlength\cmsFigWidth{0.85\columnwidth}}{\setlength\cmsFigWidth{0.4\textwidth}}
\ifthenelse{\boolean{cms@external}}{\providecommand{\cmsLeft}{top\xspace}}{\providecommand{\cmsLeft}{left\xspace}}
\ifthenelse{\boolean{cms@external}}{\providecommand{\cmsRight}{bottom\xspace}}{\providecommand{\cmsRight}{right\xspace}}
\newlength\cmsTabSkip\setlength{\cmsTabSkip}{1ex}

\cmsNoteHeader{SUS-19-013}
\title{Search for supersymmetry in proton-proton collisions at \texorpdfstring{$\sqrt{s} = 13\TeV$}{sqrt(s)=13 TeV} in events with high-momentum \PZ bosons and missing transverse momentum}

\date{\today}

\abstract{
A search for new physics in events with two highly Lorentz-boosted \PZ bosons and large missing transverse momentum is presented.
The analyzed proton-proton collision data, corresponding to an integrated luminosity of 137\fbinv, were recorded at $\sqrt{s}=13\TeV$ by the CMS experiment at the CERN LHC.
The search utilizes the substructure of jets with large radius to identify quark pairs from \PZ boson decays.
Backgrounds from standard model processes are suppressed by requirements on the jet mass and the missing transverse momentum.
No significant excess in the event yield is observed beyond the number of background events expected from the standard model.
For a simplified supersymmetric model in which the \PZ bosons arise from the decay of gluinos, an exclusion limit of 1920\GeV on the gluino mass is set at 95\% confidence level.
This is the first search for beyond-standard-model production of pairs of boosted \PZ bosons plus large missing transverse momentum.

}

\hypersetup{
pdfauthor={CMS Collaboration},
pdftitle={Search for supersymmetry in proton-proton collisions at sqrt(s) = 13 TeV in events with high-momentum Z bosons and missing transverse momentum},
pdfsubject={CMS},
pdfkeywords={CMS, physics, SUSY, LHC}}

\maketitle
\section{Introduction}
\label{sec:introduction}

The discovery of a Higgs boson in 2012 by the ATLAS and CMS experiments~\cite{Aad:2012tfa,Chatrchyan:2012ufa,Chatrchyan:2013lba} at the CERN LHC fulfilled the predicted
particle content of the standard model
(SM).
However, within the SM as a quantum field theory, the measured Higgs boson mass of around $125\gev$ presents a special challenge as the calculated mass is unstable against corrections from loop processes when the theory is extended to higher mass scales.
In the absence of extreme fine tuning~\cite{Barbieri:1987fn,Dimopoulos:1995mi,Barbieri:2009ev,Papucci:2011wy} that would precisely cancel the divergent terms, the mass value can run up to the ultraviolet cutoff of the model at the Planck scale.
This instability of the Higgs boson mass and the entire electroweak scale is known as the gauge hierarchy problem.

One widely studied extension of the SM is supersymmetry (SUSY)~\cite{Fayet:1976cr,Nilles:1983ge,Martin:1997ns}, which posits a partner for each SM particle differing in spin by one-half unit.
For example, squarks \squark and gluinos $\PSg$ are the SUSY partners of quarks and gluons, respectively.
Depending on the mass hierarchy of these new particles, they could resolve the gauge hierarchy problem by providing necessary radiative corrections to partly cancel the SM contributions.
Furthermore, in $R$-parity conserving models~\cite{Fayet:1974pd,Farrar:1978}, SUSY particles are produced in pairs, while the lightest of them is neutral, stable, and weakly interacting.
This lightest SUSY particle (LSP) provides a suitable candidate for dark matter~\cite{Farrar:1978}, which is not described in the SM.
The typical experimental signatures of pair-produced SUSY particles with $R$-parity conserving decay chains are jets, leptons, and large missing transverse momentum (\met).

\begin{figure*}[tb]
\centering
\includegraphics[width=0.50\textwidth]{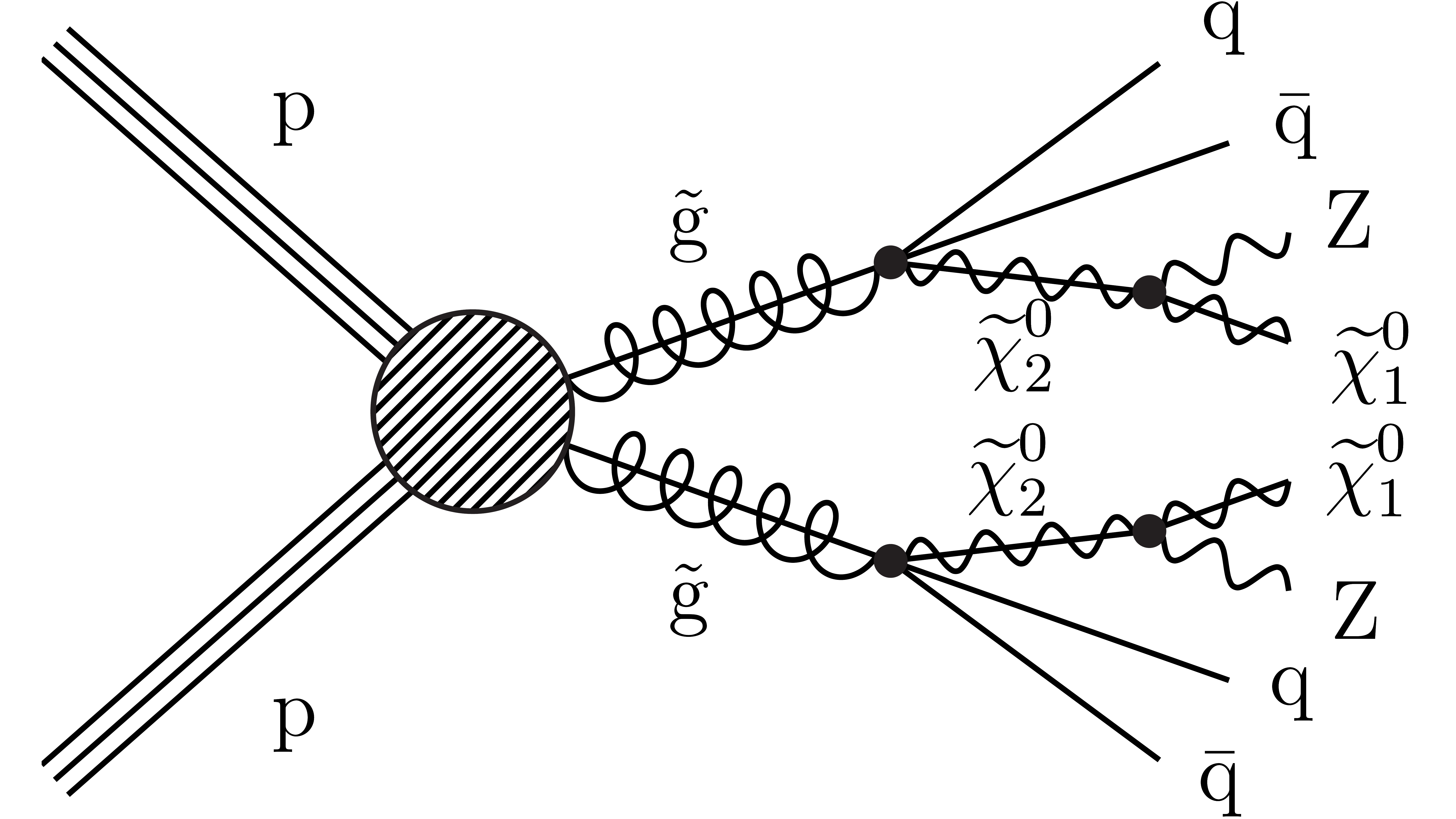}
\caption{
  Signal diagram for the T5ZZ simplified model process.  The assumed small mass splitting between the \PSg and \nlsp implies a massive \nlsp.  We further assume a 100\% branching fraction for the \nlsp decay to the \PZ boson and \lsp, leading to an energetic \PZ boson and large \met.
}
\label{fig:T5-event-diagrams}
\end{figure*}

As gluinos and squarks carry color charge, like their SM partners, they can be produced via the strong interaction;
therefore among SUSY particles they have the highest production cross sections at hadron colliders for a given mass.
Searches for direct decays of gluinos to quarks and the LSP have excluded $\mgluino\lesssim 2\TeV$~\cite{Sirunyan:2019ctn,Sirunyan:2019xwh,CMS:2019tlp,Sirunyan:2020ztc}, depending on the model.
The search described in this paper focuses on gluino decay cascades to \PZ bosons and the LSP via the next-to-lightest SUSY particle (NLSP).
We consider a picture in which the NLSP and LSP are respectively the neutralinos \NLSP and \LSP, mixed states of SUSY partners of the neutral Higgs and gauge bosons.
Such a situation arises in SUSY scenarios like those described in Ref.~\cite{PhysRevD.91.075005} that seek to preserve ``naturalness,'' that is,
minimal fine tuning of the SM to solve the gauge hierarchy problem,
by admitting large mass splittings among the neutralinos (and charginos), leading to experimental signatures with vector bosons and \met in the final state.
Figure~\ref{fig:T5-event-diagrams} shows our signal process, expressed within the framework of simplified models~\cite{bib-sms-1,bib-sms-3,bib-sms-2,bib-sms-4}, and referred to as T5ZZ.
We further assume a heavy $\NLSP$, (with mass below that of the \PSg), and a light \LSP.
This gives rise to energetic \PZ bosons along with large \met and additional soft quarks in the final state.
In our model calculations we set the branching fraction for $\NLSP\to\PZ\LSP$ to 100\%, the \LSP mass to 1\GeV, and the difference in mass between the \PSg and \NLSP to 50\GeV, though any set of mass parameters with a large [$\mathcal{O}$(TeV)] mass difference between
the \NLSP and \LSP will result in highly energetic \PZ bosons.
For the dominant $\PZ\to\qqbar$ decay at large momentum, the decay products can be contained in a single reconstructed jet with a large angular radius (wide-cone jet).

In this paper, we present a search in proton-proton ($\Pp\Pp$) collisions at
$\sqrt{s}=13\TeV$ for events with two highly Lorentz-boosted, hadronically decaying \PZ bosons and large \met.
The analysis is based on the LHC Run 2 data set with an integrated luminosity of $137\invfb$,
recorded by the CMS experiment during 2016--2018.
The signature for a signal is a pair of wide-cone jets, each having a reconstructed mass consistent with
the \PZ boson mass.  This selection, in combination with large \met,
greatly suppresses backgrounds from SM processes.

\section{The CMS detector and trigger}
\label{sec:detector}
A detailed description of the CMS detector and the associated coordinate system and kinematic variables is given in Ref.~\cite{Chatrchyan:2008aa}.
The main components of the apparatus are briefly discussed here.
The core of CMS is a cylindrical superconducting solenoid with an inner diameter of 6\unit{m} that provides a 3.8\unit{T} axial magnetic field.
A silicon pixel and strip tracker,
a lead tungstate crystal electromagnetic calorimeter,
and a brass and scintillator hadron calorimeter
are placed within the volume enclosed by the solenoid.
Gas-ionization detectors are embedded in the steel flux-return yoke outside the solenoid to identify muons.
The detector is nearly hermetic, permitting accurate measurements of~\ptmiss.

The CMS trigger system is described in Ref.~\cite{Khachatryan:2016bia}.
For this analysis, signal candidate events were recorded by requiring \ptmiss at the trigger level
to exceed a threshold that varied between 100 and 120\GeV, depending on the LHC instantaneous luminosity.
The efficiency of this trigger is measured in data to be greater
than 97\% for events satisfying the selection criteria described in Section~\ref{sec:event-selection}.
Additional triggers based on an isolated lepton or photon
are used to select control samples for
the background predictions.

\section{Simulated event samples}
\label{sec:event-samples}

The estimation of yields for the most prominent backgrounds is based on data in orthogonal signal-depleted control regions and is described in Section~\ref{sec:bkgest}.
Samples of Monte Carlo (MC) simulated events are used to test the background estimation, as well as to optimize the selection criteria.
These samples include events with top quark pair production (\ttbar), and photon, \PW boson, or \PZ boson production accompanied by jets, denoted \gjets, \wjets, or \zjets, respectively.

The SM production of \ttbar, \gjets, \wjets, \zjets, and quantum chromodynamics (QCD) multijet events is simulated using the
{\MGvATNLO}\,2.2.2~\cite{Alwall:2014hca,Alwall:2007fs} generator for 2016 samples
and {\MGvATNLO}\,2.4.2 for 2017 and 2018 samples, all with leading order (LO) precision.
The \ttbar events are generated with up to three additional partons in the matrix element calculations,
while the \gjets, \wjets, and \zjets events are generated with up to four additional partons.
Single top quark events produced via the $s$ channel, diboson events originating from
$\PW\PW$, $\cPZ\cPZ$, or $\cPZ\PH$ production,
and events from $\ttbar\PW$, $\ttbar\cPZ$, and $\PW\PW\cPZ$ production,
are generated with {\MGvATNLO}\,2.2.2 at next-to-leading order (NLO)~\cite{Frederix:2012ps},
except that $\PW\PW$ events in which both {\PW} bosons decay leptonically are generated using {\POWHEG}\,2.0~\cite{Nason:2004rx,Frixione:2007vw,Alioli:2010xd,Alioli:2009je,Re:2010bp}
at NLO. The \POWHEG generator is also used to describe $t$-channel production of
single top quarks as well as $\PQt\PW$ events.
Normalization of the simulated background samples is derived from the most accurate cross section
calculations available~\cite{Alioli:2009je,Re:2010bp,Alwall:2014hca,Melia:2011tj,Beneke:2011mq,
Cacciari:2011hy,Baernreuther:2012ws,Czakon:2012zr,Czakon:2012pz,Czakon:2013goa,
Gavin:2012sy,Gavin:2010az}, which generally correspond to NLO or next-to-NLO (NNLO) precision.

Samples of simulated signal events are generated at LO using {\MGvATNLO}\,2.2.2 (2.4.2) for the 2016 (2017 and 2018) samples,
with up to two additional partons included in the matrix element calculations. The production cross
sections are normalized to approximate NNLO plus next-to-next-to-leading logarithmic (NNLL)
precision~\cite{bib-nlo-nll-01,bib-nlo-nll-02,bib-nlo-nll-03,bib-nlo-nll-04,bib-nlo-nll-05,
Beenakker:2016lwe,Beenakker:2011sf,Beenakker:2013mva,Beenakker:2014sma,
Beenakker:1997ut,Beenakker:2010nq,Beenakker:2016gmf}.

All simulated samples make use of {\PYTHIA}\,8.205 (2016) or 8.230 (2017 and 2018)~\cite{Sjostrand:2014zea} to describe parton showering and hadronization.
The CUETP8M1~\cite{ Khachatryan:2015pea} tune was used to simulate both the SM background and signal samples for the 2016 simulation.  To generate the 2017 and 2018 samples, {\PYTHIA} was used, with the CP5 tune~\cite{Sirunyan:2019dfx} for the backgrounds and the CP2 tune~\cite{Sirunyan:2019dfx} for signals.
Simulated samples generated at LO (NLO) with the CUETP8M1 tune use the
\textsc{nnpdf3.0lo} (\textsc{nnpdf3.0nlo})~\cite{Ball:2014uwa}
 PDF set,
while those generated with the CP2 or CP5 tune use the NNPDF3.1LO
(NNPDF3.1NNLO)~\cite{Ball:2017nwa} PDF set.  Here PDF refers to the parton distribution function.
The detector response is modeled with \GEANTfour~\cite{Agostinelli:2002hh}.
The simulated events are generated with a distribution
of $\Pp\Pp$ interactions per bunch crossing (``pileup'') that is adjusted to match
the corresponding distribution measured in data.

To improve the description of initial-state radiation (ISR),
the \MGvATNLO prediction of the jet multiplicity distribution is compared with data in a control sample enriched in \ttbar events~\cite{Sirunyan:2019ctn}.  A correction factor derived therefrom
is subsequently applied to the simulated \ttbar and signal events. The correction is found to be unnecessary for \ttbar samples that are generated with the CP5 tune, so it is not applied to those samples.

\section{Event reconstruction}
\label{sec:reconstruction}

Individual particles are reconstructed
with the CMS particle-flow (PF)
algorithm~\cite{CMS-PRF-14-001},
which identifies them as photons,
charged or neutral hadrons, electrons, or muons.
These objects are characterized kinematically by their transverse momentum \pt, pseudorapidity $\eta$, and azimuthal angle $\phi$.
Photon and electron candidates are required to satisfy $\abs{\eta}<2.5$,
and muon candidates $\abs{\eta}<2.4$, within the fiducial coverage of the tracking and muon system, respectively.

The missing transverse momentum vector \ptvecmiss is computed as the negative vector sum of the \pt
of all of the PF candidates in an event, and its magnitude is denoted as \ptmiss~\cite{Sirunyan:2019kia}.
The \ptvecmiss is modified to account for corrections to the energy scale of the reconstructed jets in the event.

The reconstructed vertex with the largest value of summed physics-object
$\pt^2$ is taken to be the primary $\Pp\Pp$ interaction vertex,
where the physics objects are the jets,
clustered using the anti-\kt algorithm~\cite{Cacciari:2008gp,Cacciari:2011ma}
with the charged particle tracks assigned to the vertex as inputs,
and the associated missing transverse momentum,
taken as the negative vector \pt sum of those jets.
Charged particle tracks associated with vertices other than
the primary vertex are removed from further consideration.

Jets are defined as clusters of PF candidates
formed by the anti-\kt algorithm with a distance parameter of 0.4 or 0.8.
Quality criteria~\cite{cms-pas-jme-10-003,CMS-PAS-JME-16-003}
are imposed to suppress jets from
spurious sources such as electronics noise in the calorimeters.
The jet energies are corrected for the nonlinear response of the
detector~\cite{Khachatryan:2016kdb}.
Jets with $\pt>30\GeV$, $\abs{\eta}<2.4$, and a distance parameter of 0.4 (AK4) are used as specified in Section~\ref{sec:event-selection} to calculate some of the selection variables.
For these jets, charged particles that emerge from vertices other than the primary one are removed from the list of PF candidates used for the jet clustering.
The expected contribution from neutral particles from pileup is removed using the effective area technique~\cite{Cacciari:2007fd,CMS-PAS-JME-16-003}.

The hadronically decaying \PZ boson candidates are reconstructed as
wide-cone jets with a distance parameter of 0.8 (AK8).
These AK8 jets are reclustered from their original constituents using the ``soft drop'' method~\cite{Larkoski:2014wba} to remove soft, wide-angle radiation that can adversely impact the mass measurement of the jet.
Contributions from pileup in these jets are removed with the PUPPI technique~\cite{Bertolini:2014bba}.
The soft drop mass \mjet is then used to identify jets from $\PZ\to\qqbar$ decays.  No requirements on their flavor content are imposed.

The identification of {\PQb} jets ({\PQb} jet tagging)
is performed by applying,
to the AK4 jets,
a version of the combined secondary vertex algorithm based on
deep neural networks~\cite{Sirunyan:2017ezt} (DeepCSV).
A working point (``medium'') of this algorithm is used that has a
tagging efficiency for {\PQb} jets of 68\%,
and a misidentification probability of approximately 1\% for gluon and
light-flavor quark jets and 12\% for charm quark jets.

As described in Section~\ref{sec:event-selection}, events with leptons or photons are vetoed in the search sample selection.
Electron and muon candidates are identified as described in
Refs.~\cite{Khachatryan:2015hwa} and \cite{Sirunyan:2018fpa},
respectively.
To suppress jets erroneously identified
as leptons or genuine leptons from hadron decays,
electron and muon candidates are subjected to an
isolation requirement.
The isolation criterion is based on a variable~$\imini$,
which is the scalar \pt sum of charged hadron,
neutral hadron, and photon PF candidates within a cone
of radius $\dR=\sqrt{\smash[b]{(\Delta\eta)^2+(\Delta\phi)^2}}$
around the lepton direction,
divided by the lepton~\pt.
The expected contributions of neutral particles from
pileup
are subtracted~\cite{Cacciari:2007fd,CMS-PAS-JME-16-003}.
The radius of the cone, in radians, is 0.2 for lepton $\pt<50\GeV$,
$10\GeV/\pt$ for $50\leq\pt\leq 200\GeV$,
and 0.05 for $\pt>200\GeV$.
The decrease in cone size with increasing lepton \pt
accounts for the increased collimation of the
decay products from the lepton's parent particle
as the Lorentz boost of the latter increases~\cite{Rehermann:2010vq}.
The isolation requirement is $\imini<0.1$ (0.2) for electrons (muons).

To further suppress events with leptons from hadron decays and
single-prong hadronic $\tau$ lepton decays, the event selection veto
is extended to include
isolated charged-particle tracks not identified as electrons or muons by the criteria of the previous paragraph.
For these candidates
the scalar \pt sum of all other charged-particle tracks
within $\dR=0.3$ around the track direction,
divided by the track~\pt, is required to
be less than 0.2 if the track is identified
as a PF electron or muon,
and less than 0.1 otherwise.
Isolated tracks are required to satisfy $\abs{\eta}<2.4$.

Photon candidates are identified as described in
Ref.~\cite{Khachatryan:2015iwa}, using the ``loose'' working point,
and with an isolation requirement based on
the individual sums of energy from
charged and neutral hadrons
and electromagnetically interacting particles,
excluding the photon candidate itself,
within
$\dR=0.3$ around the direction of the photon candidate.
Each of the three individual sums,
corrected for pileup, is required not to exceed
a threshold that depends on the calorimeter geometry.

\section{Event selection}
\label{sec:event-selection}

We select events with large jet activity and \met, no leptons or photons, and wide-cone jets from Lorentz-boosted, hadronically decaying \PZ bosons. Control regions for the determination of backgrounds are also defined.

The observables used to characterize candidate events are:
\begin{itemize}
\item \njets, the number of AK4 jets in the event;
\item \met;
\item $\HT = \sum_{\text{AK4 jets}} \abs{\ptvec}$;
\item $\dphimht$, the azimuthal angle between the \ptvec of the $j^{\text{th}}$ AK4 jet and $\htvecmiss = -\sum_{\text{AK4 jets}} \ptvec$;
\item ${\mT}^i$, the transverse mass~\cite{Arnison:1983rp} of a system comprising the $i^{\text{th}}$ isolated track and \ptvecmiss;
\item \dRZb, the angular separation between a wide-cone jet and a \PQb-tagged jet.
\end{itemize}
The following requirements define the event selection:
\begin{enumerate}
\item $\njets\ge2$;
\item $\met>300 \gev$;
\item $\HT > 400 \gev$;
\item $\abs{\dphimht}>0.5\ (0.3)$ for the first two (up to next two, if $\njets >2$) AK4 jets ranked in descending order of \pt;
\item no identified isolated photon, electron, or muon candidate with $\pt>10\GeV$;
\item no isolated track with $\mT<100\GeV$ and
  \[ \pt > \left\{ \begin{array}{ll}
      5\GeV & \text{if the track is identified as a PF electron or muon},\\
      10\GeV & \text{otherwise}.
    \end{array} \right.
  \]
\item at least two AK8 jets with $\pt>200$\gev;
\item \mjet of the two highest \pt AK8 jets between 40 and 140\GeV;
\item $\dRZb > 0.8$, for the second-highest \pt AK8 jet and any \PQb-tagged jet.
\end{enumerate}
The \dphimht requirements suppress background from QCD multijet events, as well as those from hadronic $\PZ$ and $\PW$ boson decay,
for which \htvecmiss is usually aligned along a jet direction.
The \mT requirement restricts the isolated track veto to situations consistent with a {\PW} boson decay.

The first six requirements define an inclusive ``hadronic baseline'' selection, and the last three specify the further selection of
events with jet pairs that include pairs of
hadronically decaying \PZ boson candidates.
The accepted range in \mjet is chosen to reject the bulk of nonresonant SM processes
on the low side, and the peak from boosted top quark jets on the high side, while including sidebands around the \PZ boson peak to facilitate the
determination of the background.
The \dRZb requirement suppresses backgrounds from \ttbar and single
top quark events in which a top quark is reconstructed as a \PQb-tagged jet together with a \PW boson reconstructed as an AK8 jet.

\begin{figure}
  \centering
    \includegraphics[width=0.48\textwidth]{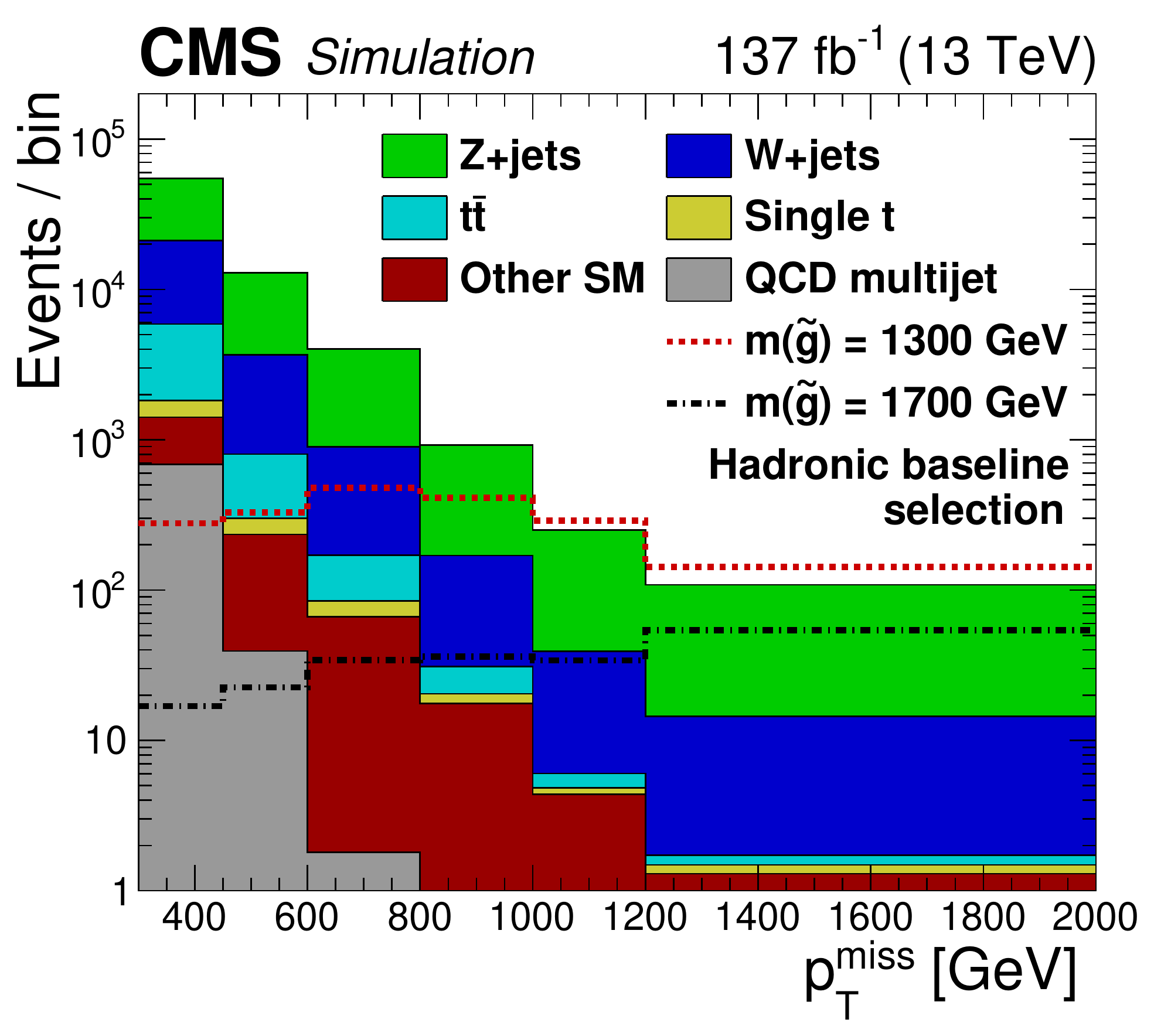}
    \includegraphics[width=0.48\textwidth]{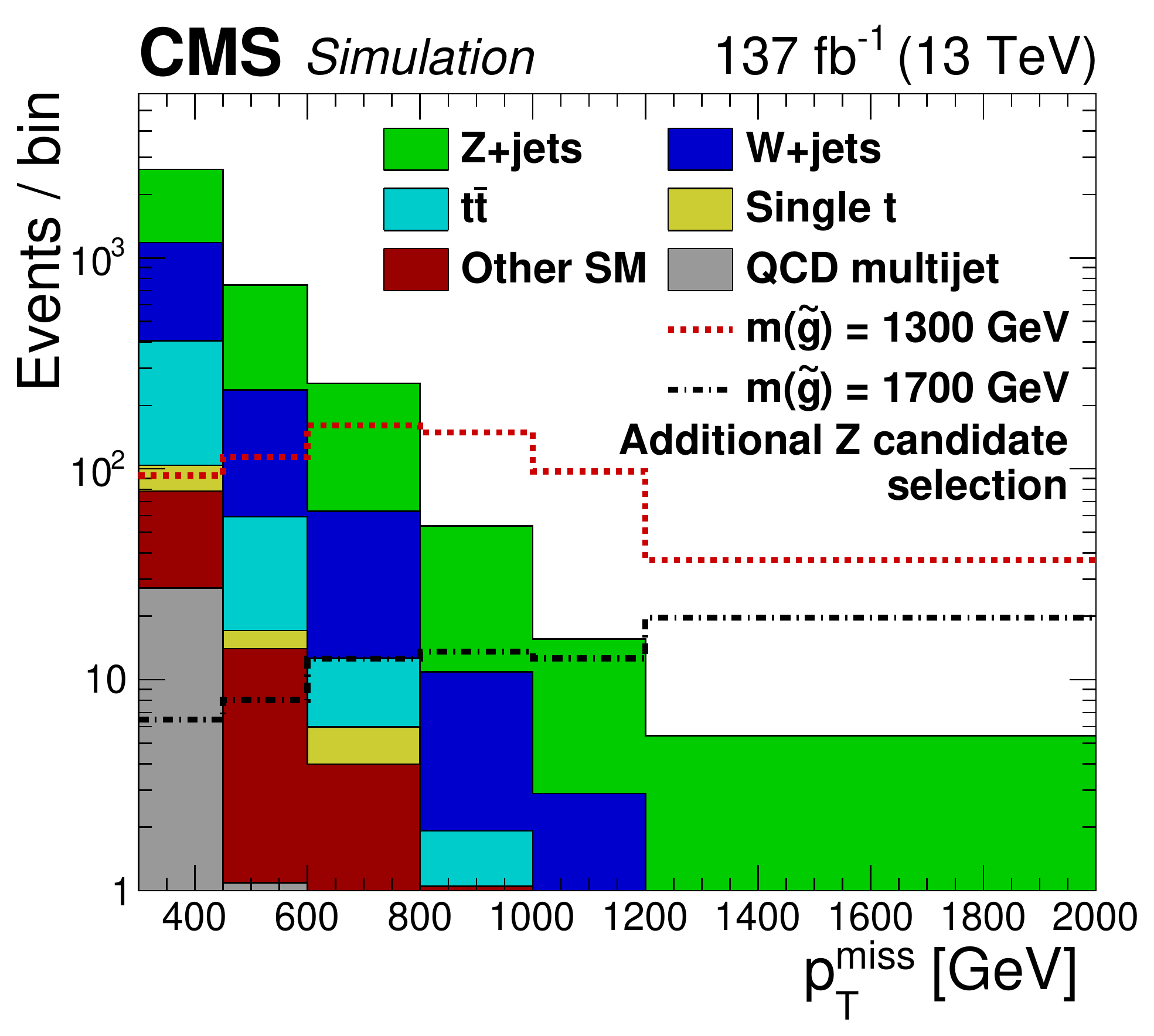}
  \caption{Distributions of \met for simulated SM backgrounds (stacked histograms), with
only the hadronic baseline selection (left), and after the additional \PZ candidate selection (right).
Expected signal contributions for two example mass points (dotted lines) are also shown.
The last bin includes the overflow events.
  }
    \label{fig:METSearchbins}
\end{figure}

Figure~\ref{fig:METSearchbins} shows the simulated SM background components and two example signal mass points for events
selected without and with the three \PZ boson requirements.
The main sources of SM background are \zjets, \wjets, and \ttbar, which can yield large \met accompanied by AK8 jets formed from random combinations of hadrons.
In the case of \zjets, large \met comes from the \znn decay. For \wjets and \ttbar, \met arises from a leptonically decaying \PW boson where the charged lepton is undetected. Smaller background contributions arise from the QCD multijet events in which the measurement of a jet's energy suffers a large fluctuation, production of single top quarks, and other SM processes, such as diboson production and \ttbar pairs accompanied by vector bosons.

An event satisfying the above criteria lies in the search region (SR) if, in addition, both of the two
highest \pt AK8 jets have \mjet values in the range [70,100]~\GeV (as discussed in Section \ref{sec:DataDrivenBkg}).
Relative to the hadronic baseline selection, about 21\% of signal events are retained
in the SR, along with 0.5\% of background events.
The \met distribution in the SR is divided into six bins, with lower boundaries at
300, 450, 600, 800, 1000, and 1200\GeV.

\section{Background estimation}
\label{sec:bkgest}
This section focuses on the estimation of SM backgrounds in each \met bin. We first describe the method based on control samples in data, then follow
with a description of
the performance of the method in simulation (MC closure), and lastly deal with the uncertainty in the \met dependence (shape uncertainty) based on the data observed in the validation samples.

\begin{figure}
  \centering
  \includegraphics[width=0.52\textwidth]{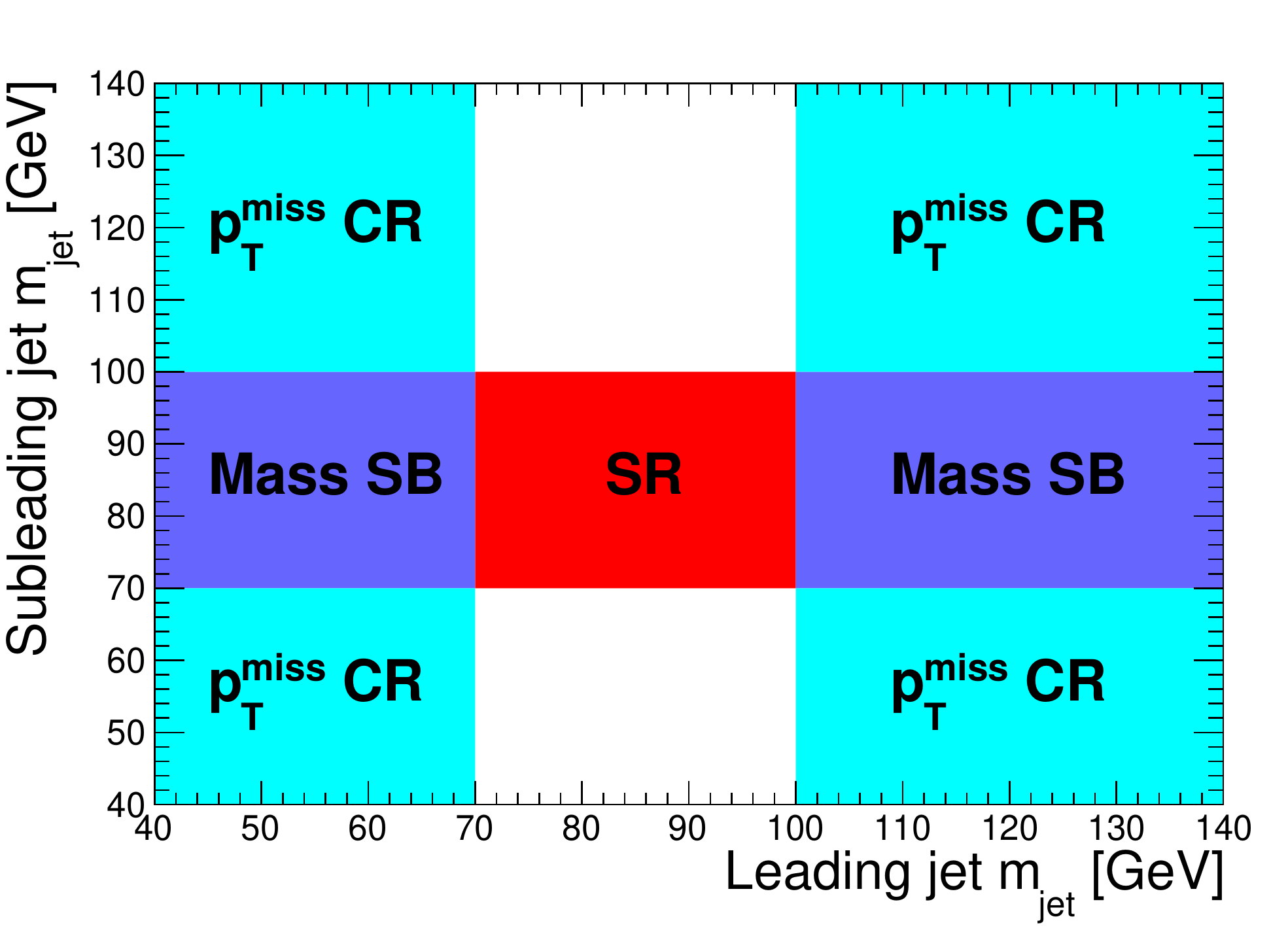}
  \caption{Definition of the search and control regions in the plane of subleading vs. leading jet mass.
    The search region (red central box), with both \mjet values lying within the \PZ signal window,
    defines the acceptance for potential signal;
    the leading-jet mass sideband (dark blue), with subleading jet within and leading jet
    outside the signal window, is used to measure the background normalization;
    the \met CR (light blue), with both leading- and subleading-jet \mjet values lying outside
    the signal window, is used to derive the \met shape in the search region.}
\label{fig:SignalControl}
\end{figure}

\subsection{Background estimation method}
\label{sec:DataDrivenBkg}

Control regions (CRs) are formed from the events in which one or both of the
highest \pt (leading) and second-highest \pt (subleading) jets lie in the \mjet sideband
$\left[40,70\right]\lor\left[100,140\right]\GeV$.
Figure~\ref{fig:SignalControl} shows the definition of the SR and CRs in the plane of jet masses of the leading and subleading jets.
In addition, validation samples are selected by inverting the lepton or photon veto requirement.

\begin{figure}[htbp]
  \centering
    \includegraphics[width=0.52\textwidth]{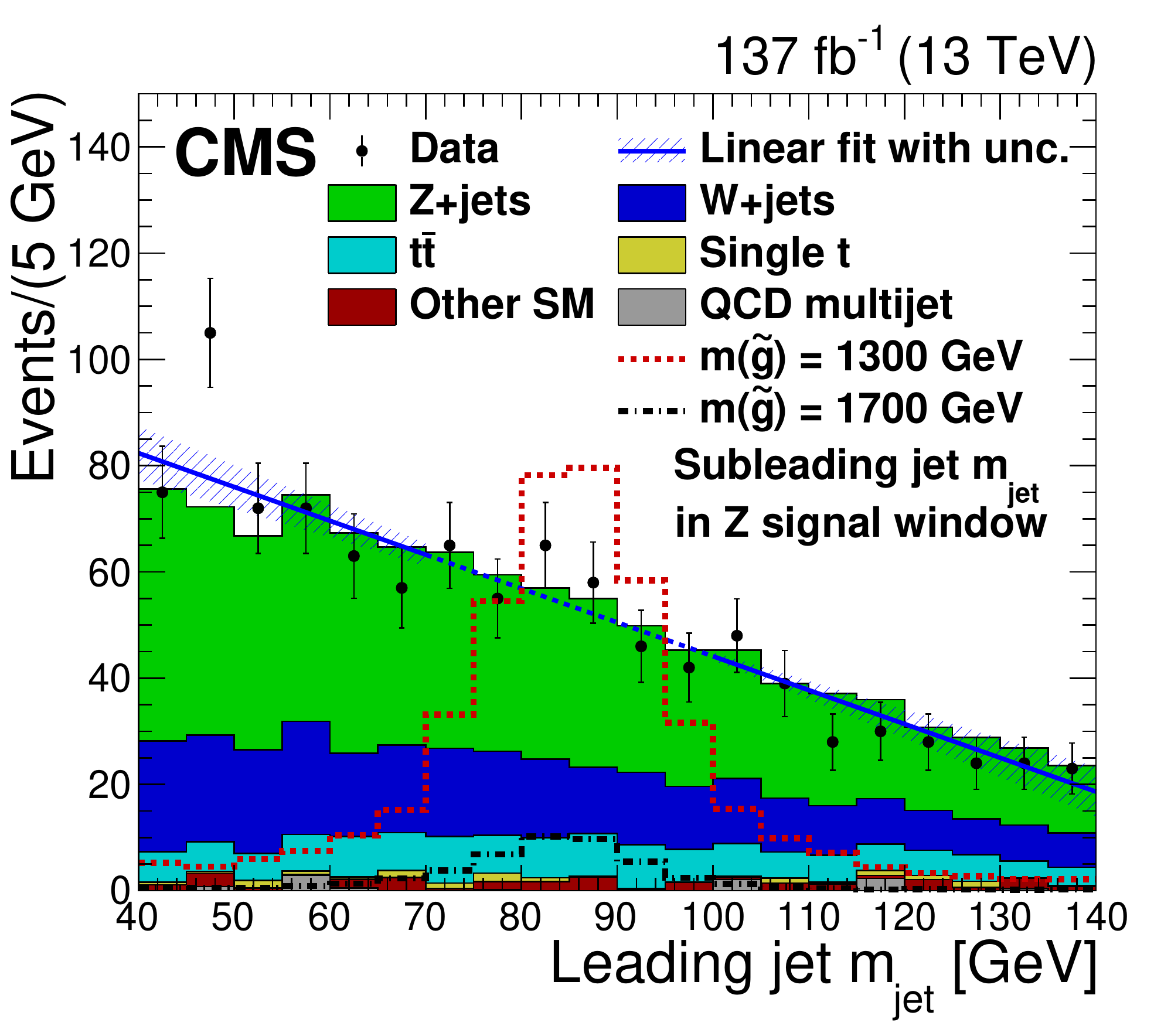}
    \caption{Leading AK8 jet \mjet shape fit in the mass sidebands. The \PZ candidate selection is applied and the subleading AK8 jet \mjet value is required to lie in the \PZ signal window. The blue hatched region represents the $\pm 1$ standard deviation uncertainty in the fit to the mass sideband performed with a linear function, which is indicated by the blue line. The stacked histogram shows the background from simulation scaled to the data.
Expected signal contributions for two example mass points are also shown.
 }
    \label{fig:Databkgmassfitting}
\end{figure}

The first step of the method is to determine the background normalization
\Bnorm integrated over all \met\ bins above $300\GeV$.
We fit the \mjet distribution for the leading jet in the leading-jet mass sideband, defined as the sample having the subleading jet \mjet within, and the leading jet \mjet outside, the \PZ signal window.
The bulk of the
background is from nonresonant SM contributions, which can be modeled with a smoothly falling shape.
The nominal fit is performed with a linear function, as shown in
Fig.~\ref{fig:Databkgmassfitting}.

The uncertainties in \Bnorm include a statistical component from the fit,
and a systematic one due to the choice of the fitting function.
To obtain the statistical uncertainty due to the interpolation of the fit into the SR, pseudo-experiments generated from the background model are fitted using a linear function with free slope and normalization.
The Gaussian width of the resulting distribution of the yields in the \PZ signal window, 10.7 events, is taken as
the statistical uncertainty in the total background prediction.

To test if the linear function is adequate to represent the \mjet distribution, we consider higher-order polynomials as alternative functions.
We check Chebyshev polynomials of up to the fourth order.
The largest variation in the fitted yield with respect to the nominal one,
10.9 events, comes from a fit with a third-order Chebyshev polynomial, and is taken as an additional uncertainty attributable to the fit shape.
Considering the statistical uncertainty described above, this results in $\Bnorm = 325 \pm 15$.

To determine the distribution of background events in the \met bins, we rely on an
underlying assumption that \met and \mjet have minimal correlation. To derive the \met shape in the SR, a nonoverlapping CR is used in which both leading and subleading AK8 jets have \mjet in the mass sideband.
This is referred to as the \met CR (Fig.~\ref{fig:SignalControl}). In each of the six \met bins, we
calculate the background prediction as
\begin{linenomath}
  \begin{equation}
    \label{eq:bkeq}
          \mathcal{B}_i = \mathcal{T}N^{\mathrm{CR}}_i,
  \end{equation}
\end{linenomath}
where $N^{\mathrm{CR}}_i$ is the yield in \met bin $i$ in the \met CR, and the transfer factor,
\begin{linenomath}    
\begin{equation}
  \label{eq:tf}
  {\mathcal{T}} \equiv \frac{\Bnorm}{\sum_i{N^{\mathrm{CR}}_i}} = 0.198\pm0.009,
\end{equation}
\end{linenomath}
scales the \met CR yield to that of the SR.
The uncertainty in ${\mathcal{T}}$ includes both statistical and systematic uncertainties in $\Bnorm$.

\subsection{Background closure in simulation}
\label{sec:bkgclosure}

The background estimation method based on control samples in data is tested by applying the procedure to MC simulation. We perform this closure test in two steps.

\begin{figure}[htbp]
  \centering
    \includegraphics[width=0.48\textwidth]{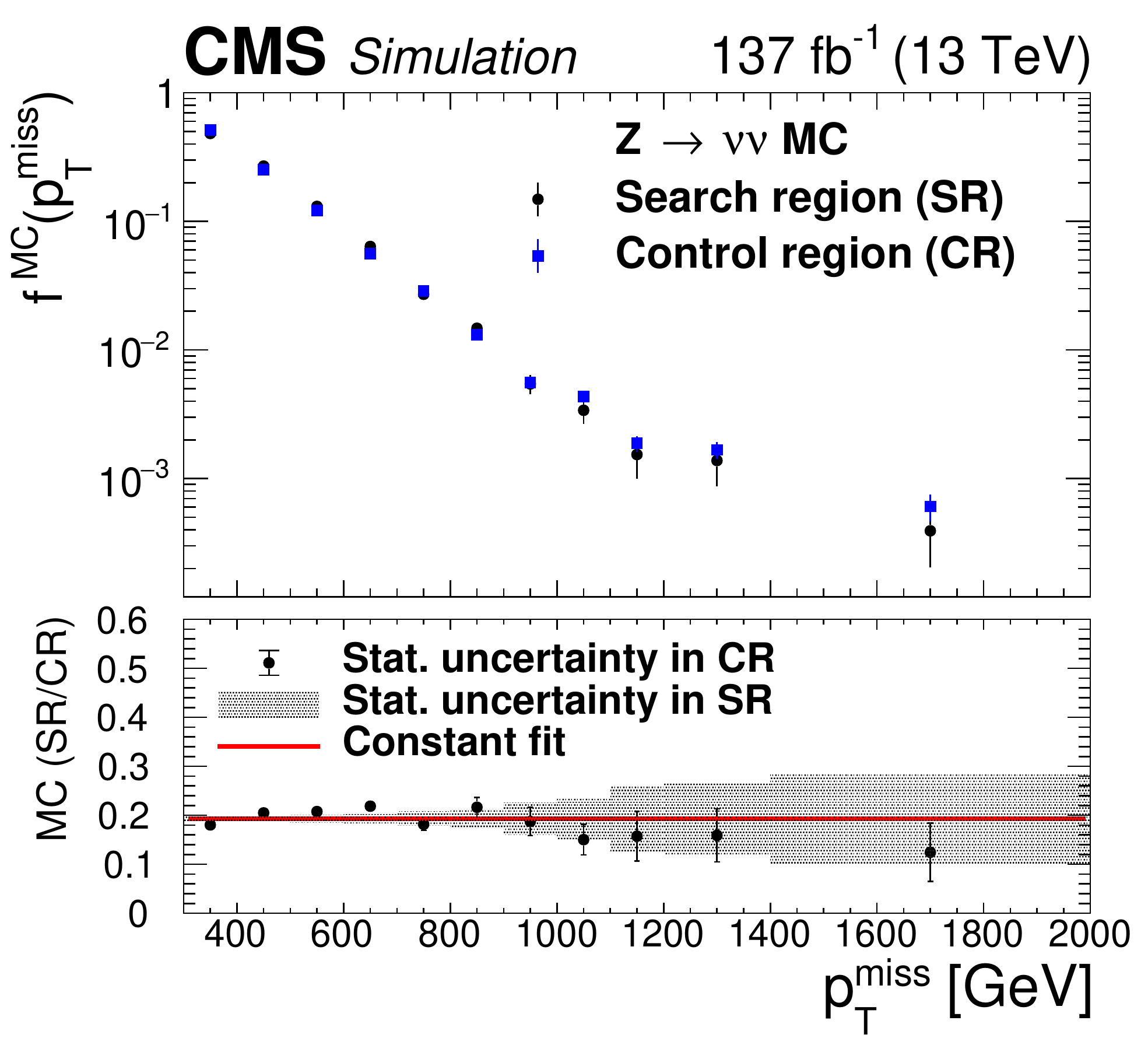}
    \includegraphics[width=0.48\textwidth]{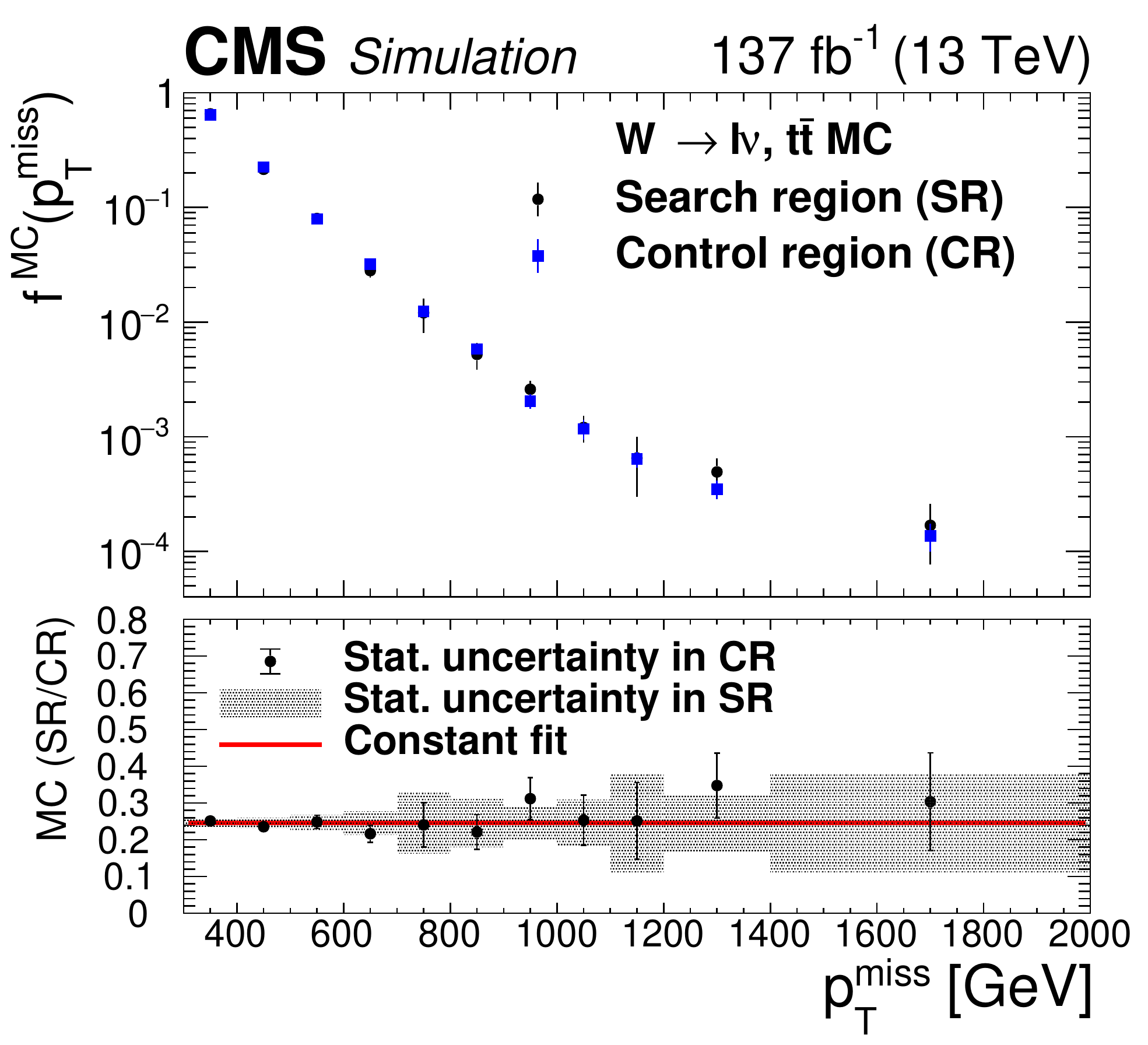}
\caption{Comparison of the \met shape in the search and control regions in simulation. The upper panels show the unit-normalized \met distributions $f^{\text{MC}}(\met)$ in the two regions, while the lower panels
  show the ratio of the number of events in the search region to that in the control region. This comparison is done for two main background components: \znn (left) and \ttbar plus \wjets (right).
In the lower panel the statistical uncertainties in the search and control region yields are denoted by the shading and vertical bars, respectively, and
a fit to a constant is included to show the average ratio.}
\label{fig:MCSRValidation}
\end{figure}

The main assumption to verify is the lack of correlation between the AK8 jet mass and \met shape.
Figure~\ref{fig:MCSRValidation} shows the results of a test of this assumption,
where the simulated sample size permits a distribution in relatively fine steps.
The plots compare the \met shape in the search and control regions,
for the two main background processes.
In both cases we see that the \met shapes are consistent between the two regions.

\begin{figure}[htbp]
  \centering
    \includegraphics[width=0.6\textwidth]{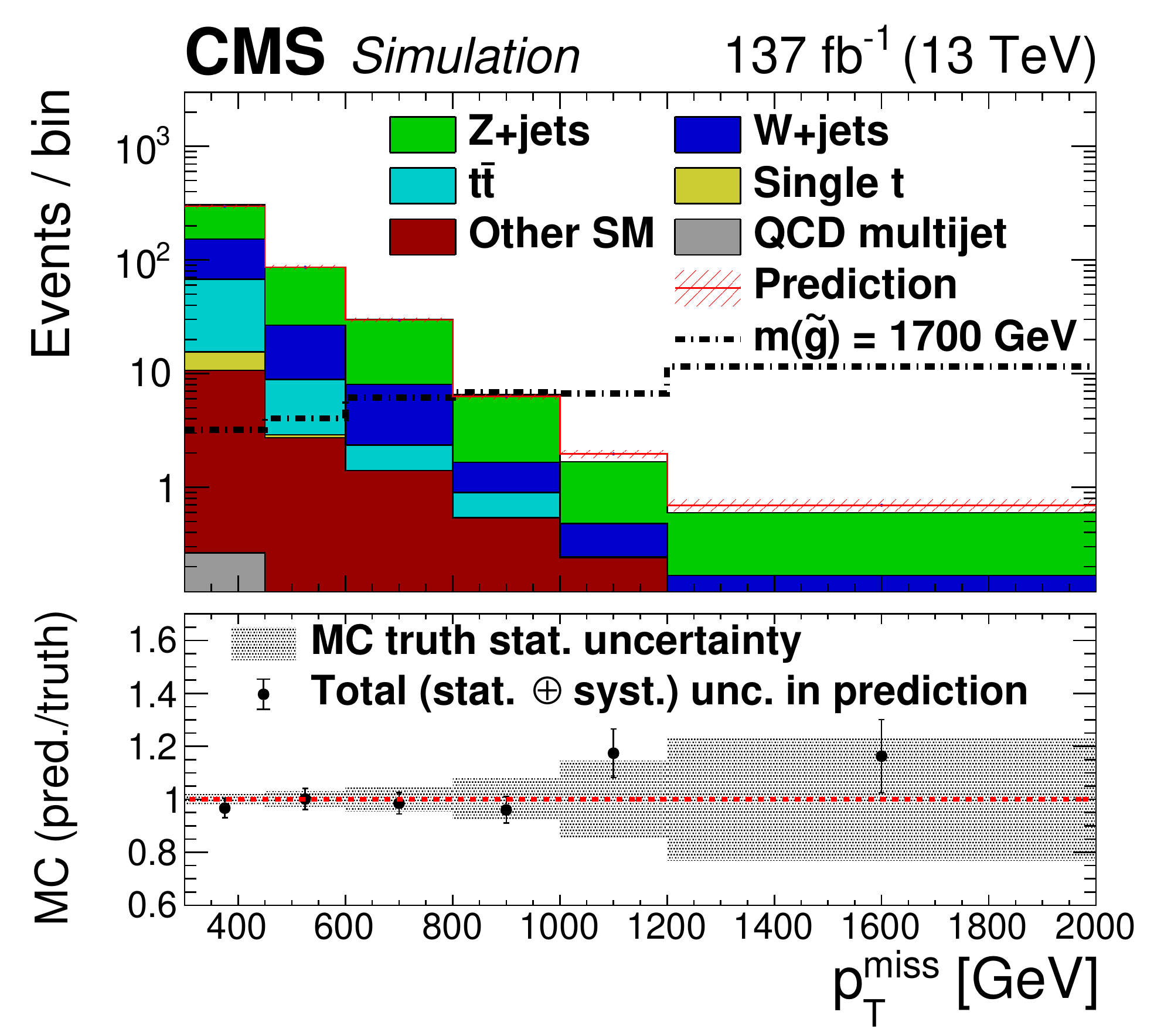}
\caption{
Results of the closure test in which the background estimation method based on control samples in data is applied to simulation and compared with the direct yield, in the analysis search bins.
Expected signal contribution for one example mass point is also shown.
The lower panel shows the ratio of the prediction to the direct yield.  The gray band shows the statistical uncertainty in the direct yield, and the error bars on the points
represent the total uncertainty in the prediction.}
\label{fig:MCClosure}
\end{figure}

For the closure test of the background estimation method we calculate the
background prediction in each \met bin
[Eq.~(\ref{eq:bkeq})] and compare these predictions with the background yields taken directly from simulation.
The results of this test, shown in Fig.~\ref{fig:MCClosure}, demonstrate
good agreement within the statistical precision of the test.
To account for the uncertainties in the comparison, we assign
the relative difference between the prediction and direct observation as a nonclosure systematic uncertainty in the \met shape.
This difference ranges from 1 to 20\%, where the variations in the four lower \met bins are treated as being anti-correlated
with those in the higher \met bins to give a systematic uncertainty in the \met shape that does not affect the overall normalization of the background estimation.

\subsection{The \texorpdfstring{\met}{Missing transverse momentum} shape uncertainty}
\label{sec:validation}

While the background estimation method is shown to close well in simulation, we additionally verify in data how well the \met CR models the \met shape in the \PZ signal window.
In particular, two validation samples are used to compare the \met\ shape obtained from the \met CR with the one obtained in the \PZ signal window, used to define our
SR, for the main background components. A photon validation sample is used as a proxy for the \zjets background component, while a
single-lepton sample is used to validate the modeling of \ttbar and \wjets combined.

We select the photon validation sample from events recorded with a
single-photon trigger, replacing the photon veto with the requirement of exactly one photon, defined as in Section \ref{sec:reconstruction}. The photon \pt is used to emulate the \met from the \PZ boson when the latter decays to neutrinos. The lower-\pt trigger
threshold for the photon compared with the \met threshold in the signal trigger allows us to consider the photon validation sample down to
$200\gev$ in photon \pt as a proxy for \met.
To enhance the event count in this sample, we do not require a threshold on \dRZb since there is a low risk
of heavy flavor contamination.
All other event selection requirements are the same as for the SR of the analysis.

For the single-lepton sample, the same \met\ trigger is used as for the SR. The same offline criteria are also applied,
with the exception that the \met\ requirement is relaxed to $200\gev$ to gain a longer lever arm for the \met\ shape comparison,
and the lepton vetoes are applied only after selecting exactly one electron or muon.

\begin{figure}[htbp]
\centering
\includegraphics[width=0.48\textwidth]{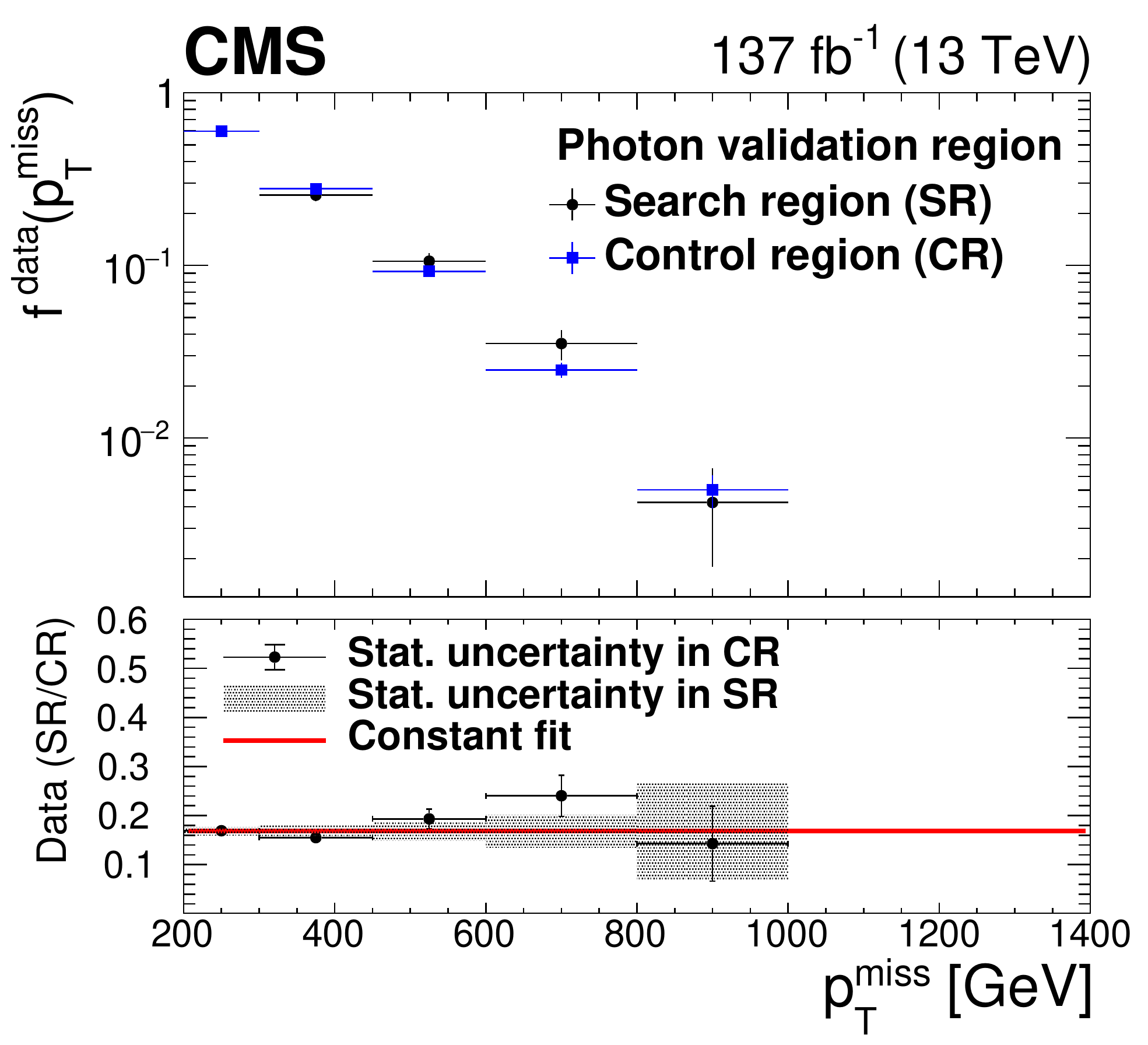}
\includegraphics[width=0.48\textwidth]{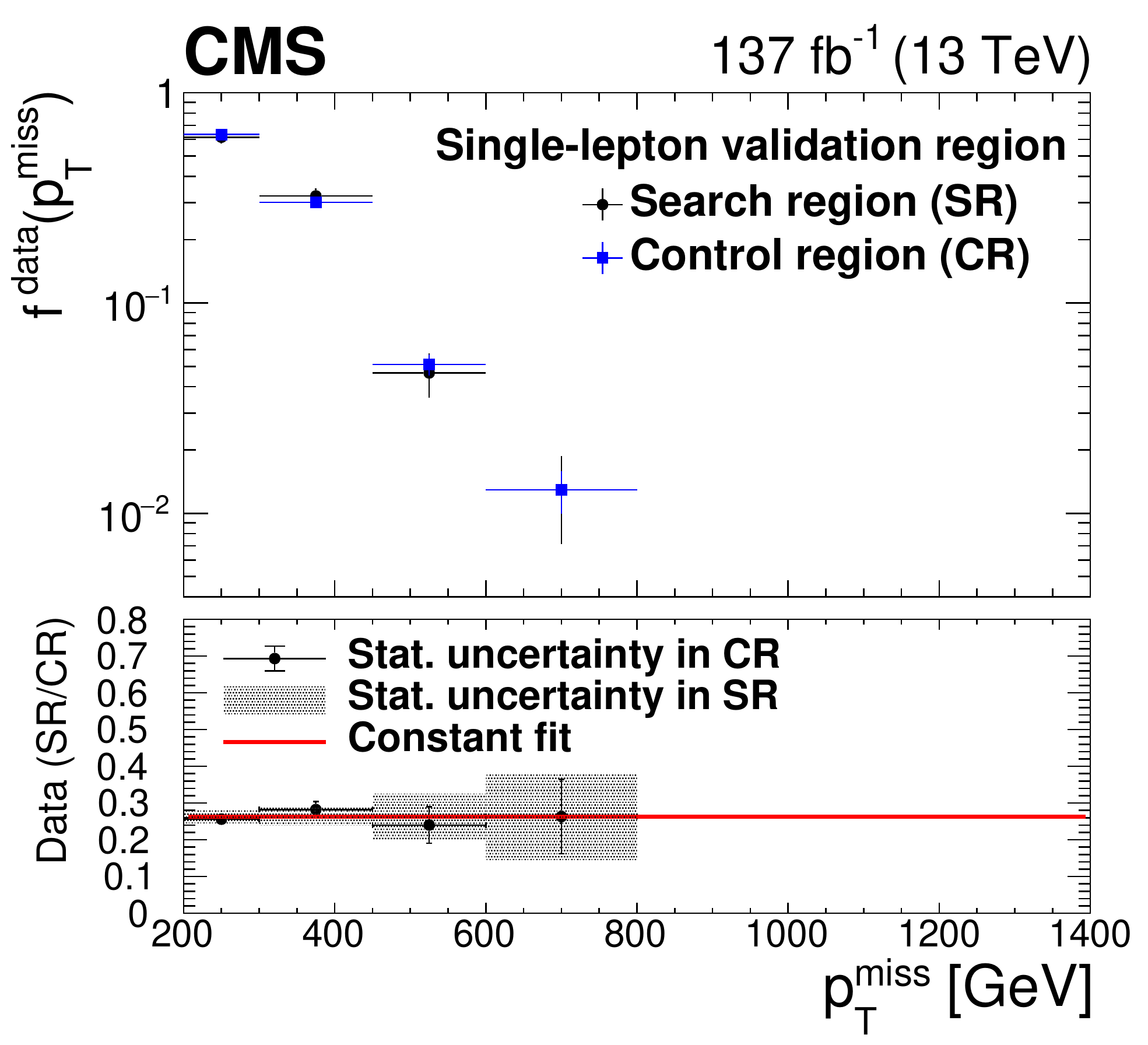}
\caption{Comparison of the \met shape between the \PZ signal window and \met control region for the photon (left)
and single-lepton (right) validation samples in data. The upper panels show the unit-normalized \met distributions $f^{\text{data}}(\met)$ in the two regions, while the lower panels
show the ratio of the number of events in the search region to that in the control region. A fit to a constant is included in the lower panels to show the average ratio. The horizontal bars on the markers indicate the widths of the search bins.
In the lower panel the statistical uncertainties in the search and control region yields are denoted by the shading and vertical bars, respectively.}
\label{fig:ValidationData}
\end{figure}

Figure~\ref{fig:ValidationData} shows the \met shape comparison for the photon and single-lepton data. Both ratios
are consistent with being independent of \met, as expected from the MC closure test, albeit within the limited statistical precision of the data.
To account for possible shape differences between the search and control regions, we apply a
systematic uncertainty in the \met shape calculated using the photon and single-lepton samples.
The uncertainty is the difference with respect to a uniform distribution of a fit to the SR/CR distribution with a
linear function having a free slope parameter.
This results in uncertainties ranging from 0--33\% in the \zjets background based on the photon validation sample, and 1--14\% in the combined \ttbar and \wjets background based on the single-lepton validation sample.  Weighting these by the proportions of those components in the total background yields uncertainties of 2--30\%, depending on the \met bin.

\section{Systematic uncertainties}
\label{sec:signalsys}

\begin{table}[h!t]
\topcaption{
    Summary of systematic uncertainties, where the ranges refer to different \met bins. In the last column we distinguish uncertainties that affect the normalizations ("norm."), the shapes of distributions, or both.
}
\centering
\begin{tabular}{lcc}
\hline
Source of uncertainty & Effect on yields (\%) & norm. or shape \\
\hline
\multicolumn{3}{c}{Uncertainties in the background predictions} \\
Fit, normalization & 3.3 & norm. \\
Fit, shape & 3.4 & norm. \\
\mjet CR statistics & 3--100 & shape \\
MC closure & 2--13 & shape \\
Data validation & 2--30 & shape \\ [\cmsTabSkip]
\multicolumn{3}{c}{Uncertainties in the signal yields} \\
Integrated luminosity & 2.3--2.5 & norm.\\
Trigger efficiency & 2.0 & both \\
Isolated lepton and track vetoes & 2.0 & norm. \\
Jet quality requirements & 1.0 & norm. \\
ISR modeling & 1--2 & both \\
\rscale and \fscale scales & 0.2--0.5 & both \\
JEC & 2--4 & both \\
JER & 5--6 & both \\
MC statistics & 1--2 & both \\
\mjet resolution & 1--3 & norm. \\
\hline
\end{tabular}
\label{tab:SystSummary}
\end{table}

The uncertainties in the SM background prediction are described in Section~\ref{sec:bkgest},
along with the description of the background estimation method.
The uncertainties in the background normalization include the statistical uncertainty
from the mass sideband fit interpolation as well as the systematic one derived from alternative fit functions.
The uncertainties in the \met shape include the statistical uncertainties of the \met CR. The systematic uncertainties only affect the \met shape
without changing the background normalization. These are derived from the MC closure test and data validation samples.
All of these systematic uncertainties are summarized in the upper section of Table~\ref{tab:SystSummary}.

The sources of uncertainty in the signal efficiency affect the signal normalization,
the signal \met shape, or both, as indicated in Table~\ref{tab:SystSummary}.
The uncertainties in the integrated
luminosity are 2.5\%~\cite{CMS-PAS-LUM-17-001},
2.3\%~\cite{CMS-PAS-LUM-17-004}, and 2.5\%~\cite{CMS-PAS-LUM-18-002}
for 2016, 2017, and 2018, respectively.
The trigger, lepton veto, and isolated-track veto efficiencies are
measured in data validation samples and their statistical uncertainties propagated to the signal yields.
The ISR modeling in the simulation is adjusted to match the efficiencies measured in data events enriched in dileptonic \ttbar production and decay,
and the uncertainty in this correction is propagated to the signal yields.
To evaluate the uncertainty associated with the
renormalization (\rscale) and factorization (\fscale) scales,
each scale is varied independently
by a factor of 2.0 and 0.5~\cite{Catani:2003zt,Cacciari:2003fi}.
Uncertainties in the simulation of pileup are found to be of the order of $0.02\%$; thus no associated uncertainty is applied.

The jet momenta in MC samples are smeared to match the jet energy resolution (JER) in data.
The jet energy corrections (JECs) are varied using \pt- and $\eta$-dependent uncertainties.
Both effects are propagated to the jet-dependent variables, including \met, \HT, and \dphimht, and are varied within the uncertainty of the corrections to derive a systematic uncertainty in the signal yields.
The efficiency of the jet quality requirements used to suppress events with misreconstructed jets is found to differ by 1\% between data and simulation, and this is applied as
a systematic uncertainty.
The difference in the resolution of \mjet between data and simulation is applied as a smearing
factor to the MC events, and the statistical uncertainty in the size of the correction is included as a
systematic uncertainty in the corresponding selection efficiency.
Lastly, the statistical precision due to the limited event count in the simulated samples
is accounted for as an uncertainty.

The systematic uncertainties associated with the signal yields are evaluated assuming that the contributions from the three years of data taking are fully correlated.
The total systematic uncertainties in the signal yields range from 0.2 to 6\%.

\section{Results}
\label{sec:results}

\begin{figure}[htbp!]
  \centering
    \includegraphics[width=0.6\textwidth]{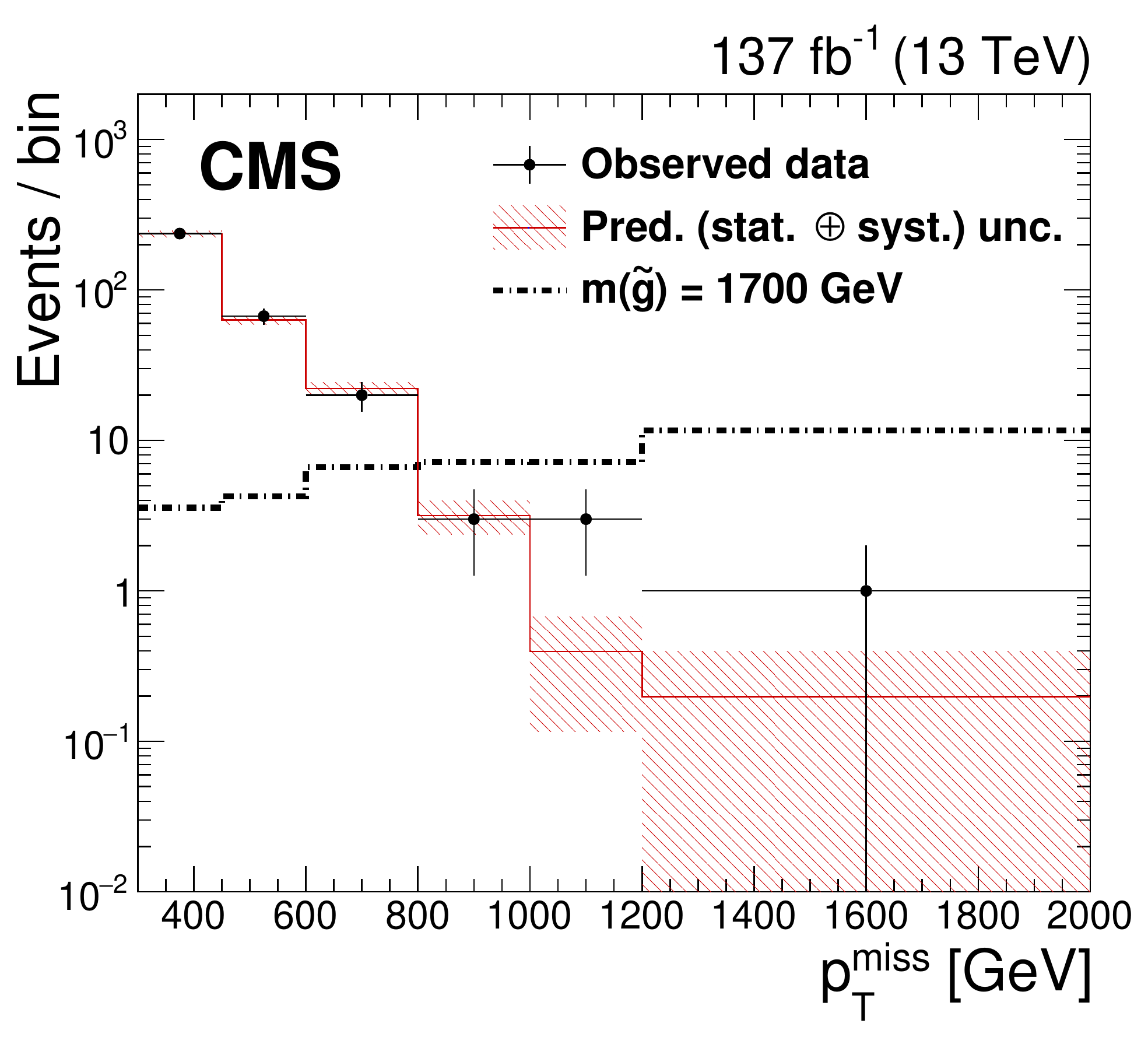}
    \caption{
    Observed data and background prediction as functions of \met. The horizontal bar associated with each data point represents the width of the corresponding bin. The red hatched region denotes the expected statistical and systematic uncertainties added in quadrature.
Expected signal contribution for one example mass point is also shown.
    }
    \label{fig:results}
\end{figure}

The background predictions and observed yields for each \met bin are shown in
Fig.~\ref{fig:results} and Table~\ref{tab:results}.  The table also gives the inputs to the prediction calculation, Eq.~(\ref{eq:bkeq}).  The observations are
found to be consistent with the SM predictions within uncertainties,
and no evidence for SUSY is observed.
We calculate upper limits on the gluino pair-production cross section using a maximum-likelihood fit
in which the free parameters are
the signal strength $\mu$ and the nuisance parameters associated with the systematic uncertainties
in the background and signal model.
The uncertainty in the normalization of the background is represented with a lognormal function correlated across all \met bins, while
the \met CR statistical uncertainties are assigned as uncorrelated. The MC closure and data-MC agreement uncertainties are assigned
as correlated across \met bins.

We evaluate 95\% confidence level (\CL) upper limits based on the asymptotic form of a likelihood ratio test
statistic~\cite{Cowan:2010js}, in
conjunction with the \CLs criterion described in Refs.~\cite{bib-cls,Junk1999,cms-note-2011-005}.  The test statistic is
$q(\mu) = -2 \ln(\mathcal{L}_{\mu}/\mathcal{L}_{\text{max}})$, where $\mathcal{L}_{\mu}$ is the maximum likelihood for fixed $\mu$, and
$\mathcal{L}_{\text{max}}$ is the same determined by allowing all parameters, including $\mu$, to vary.

\begin{figure}[htbp!]
  \centering
    \includegraphics[width=0.8\textwidth]{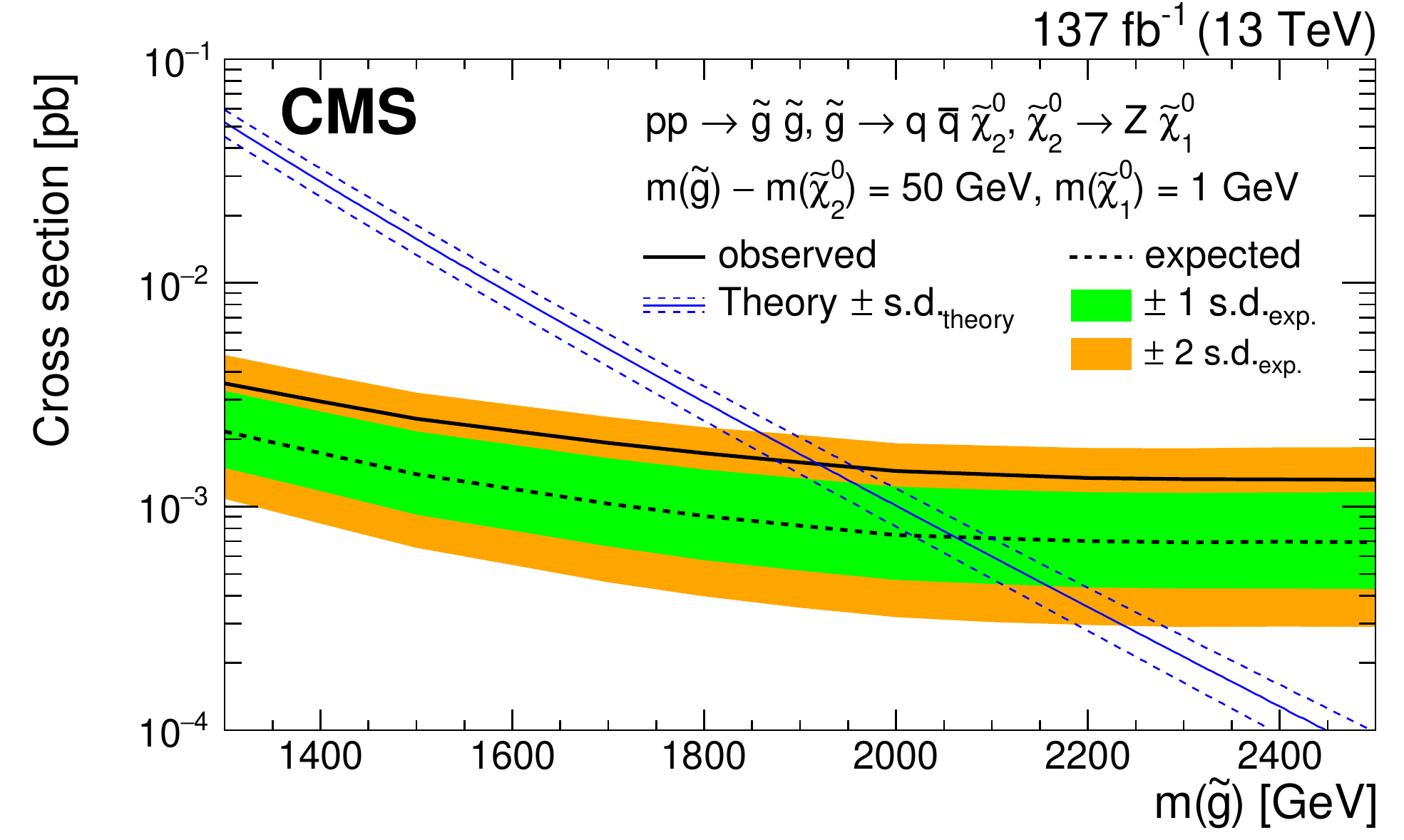}
    \caption{The
    95\% \CL upper limit on the production cross section for the T5ZZ signal model
    as a function of the gluino mass.
      The solid black curve shows the observed exclusion limit.
      The dashed black curve presents the expected limit
      while the green and yellow bands represent the $\pm 1$ and $\pm 2$ standard
      deviation
      uncertainty ranges.
      The approximate-NNLO+NNLL cross
      sections~\cite{bib-nlo-nll-01,bib-nlo-nll-02,bib-nlo-nll-03,
      bib-nlo-nll-04,bib-nlo-nll-05}
      are shown in the solid blue curve while the dashed blue curves show their theoretical uncertainties~\cite{Borschensky:2014cia}. The T5ZZ model assumes a 100$\%$ branching fraction for the \NLSP to decay to the \PZ boson and \LSP.
    }
    \label{fig:LimitsT5ZZ}
\end{figure}

Expected and observed 95\% \CL upper limits, and the predicted gluino pair-production cross sections, are shown in
Fig.~\ref{fig:LimitsT5ZZ},
taking $\mlsp=1\GeV$ and $\mgluino-m(\PSGczDt)=50\GeV$.
The observed (expected) gluino mass limits reach as high as $1920$\ $(2060)~\GeV$.
The observed limit is 1.4 standard deviations weaker than the expected one due to the mild
excesses observed in the two highest \met\ bins.
The sensitivity of the search is independent of $m(\LSP)$ values that are small compared with $m(\NLSP)$, and of $m(\NLSP)$ values large enough to ensure Lorentz-boosted \PZ boson daughters.  A gradual loss of signal efficiency occurs with increasing $\Delta m(\PSg, \NLSP)$ as quarks from the gluino decay that form AK8 jets with \pt above the 200\GeV threshold displace \PZ jets as leading or subleading in \pt.

\begin{table}[htbp!]
\caption{Number of events in the \met CR, transfer factor, background prediction, and observed yield in each of the six \met bins.
Where two uncertainties are quoted, the first is statistical and the second systematic.  The systematic uncertainties in the background prediction include the shape uncertainties in addition to the uncertainty in $\mathcal{T}$.
Also listed in the last column is the number of expected signal events and corresponding statistical uncertainties for one example mass point.}
\centering
\begin{tabular}{cccccc}
\hline
\met bin    & \met CR   & Transfer                  & Background              & Observed        & Exp. signal \\
(\GeV{})&yield $N^{\mathrm{CR}}$& factor $\mathcal{T}$  & prediction $\mathcal{B}$& yield          & \mgluino = 1700 \GeV \\
            &  (events) &	                    & (events)                & (events)       & (events) \\
\hline
300--450   & 1191 &\multirow{6}{*}{$0.198\pm0.009$} & $236\pm 7 \pm 16$      & 237 & $3.5 \pm 0.1$     \\
450--600   & 320  &                                & $63.3\pm 3.6 \pm 3.3$   & 67  & $4.3 \pm 0.1$     \\
600--800   & 112  &                                & $22.2\pm 2.0 \pm 1.9$   & 20  & $6.6 \pm 0.1$     \\
800--1000  & 16   &                                & $3.2 \pm 0.8 \pm 0.5$   & 3   & $7.2 \pm 0.1$     \\
1000--1200 & 2    &                                & $0.40\pm 0.29 \pm 0.11$ & 3   & $7.2 \pm 0.1$     \\
$>$1200    & 1    &                                & $0.20\pm 0.20 \pm 0.06$ & 1   & $11.6 \pm 0.1$     \\
\hline
\end{tabular}
\label{tab:results}
\end{table}

\section{Summary}
\label{sec:summary}

Results are presented of a search for events with two hadronically decaying, highly energetic \PZ bosons and large transverse
momentum imbalance, in proton-proton collisions at $\sqrt{s} = 13\TeV$.
The sample corresponds to an integrated luminosity of 137\fbinv.
The signature for a \PZ boson candidate is a wide-cone jet having a measured mass compatible with
the \PZ boson mass.  Yields from standard model background processes, which are small for events with the largest transverse momentum imbalance, are estimated from the data in jet mass sidebands.
No evidence for physics beyond the standard model is observed.
The reach of the search is interpreted in a simplified supersymmetric model of gluino pair production in which each gluino decays to a low-momentum quark pair
and the next-to-lightest supersymmetric particle (NLSP), and the latter decays to a \PZ boson and the lightest supersymmetric particle (LSP).
With the further assumption of a large mass splitting between the NLSP and LSP, the data exclude gluino masses below 1920~\GeV at 95\% confidence level.
This is the first search for beyond-standard-model production of pairs of boosted \PZ bosons plus large missing transverse momentum.

\begin{acknowledgments}
  We congratulate our colleagues in the CERN accelerator departments for the excellent performance of the LHC and thank the technical and administrative staffs at CERN and at other CMS institutes for their contributions to the success of the CMS effort. In addition, we gratefully acknowledge the computing centers and personnel of the Worldwide LHC Computing Grid for delivering so effectively the computing infrastructure essential to our analyses. Finally, we acknowledge the enduring support for the construction and operation of the LHC and the CMS detector provided by the following funding agencies: BMBWF and FWF (Austria); FNRS and FWO (Belgium); CNPq, CAPES, FAPERJ, FAPERGS, and FAPESP (Brazil); MES (Bulgaria); CERN; CAS, MoST, and NSFC (China); COLCIENCIAS (Colombia); MSES and CSF (Croatia); RIF (Cyprus); SENESCYT (Ecuador); MoER, ERC IUT, PUT and ERDF (Estonia); Academy of Finland, MEC, and HIP (Finland); CEA and CNRS/IN2P3 (France); BMBF, DFG, and HGF (Germany); GSRT (Greece); NKFIA (Hungary); DAE and DST (India); IPM (Iran); SFI (Ireland); INFN (Italy); MSIP and NRF (Republic of Korea); MES (Latvia); LAS (Lithuania); MOE and UM (Malaysia); BUAP, CINVESTAV, CONACYT, LNS, SEP, and UASLP-FAI (Mexico); MOS (Montenegro); MBIE (New Zealand); PAEC (Pakistan); MSHE and NSC (Poland); FCT (Portugal); JINR (Dubna); MON, RosAtom, RAS, RFBR, and NRC KI (Russia); MESTD (Serbia); SEIDI, CPAN, PCTI, and FEDER (Spain); MOSTR (Sri Lanka); Swiss Funding Agencies (Switzerland); MST (Taipei); ThEPCenter, IPST, STAR, and NSTDA (Thailand); TUBITAK and TAEK (Turkey); NASU (Ukraine); STFC (United Kingdom); DOE and NSF (USA).
    
  \hyphenation{Rachada-pisek} Individuals have received support from the Marie-Curie program and the European Research Council and Horizon 2020 Grant, contract Nos.\ 675440, 752730, and 765710 (European Union); the Leventis Foundation; the A.P.\ Sloan Foundation; the Alexander von Humboldt Foundation; the Belgian Federal Science Policy Office; the Fonds pour la Formation \`a la Recherche dans l'Industrie et dans l'Agriculture (FRIA-Belgium); the Agentschap voor Innovatie door Wetenschap en Technologie (IWT-Belgium); the F.R.S.-FNRS and FWO (Belgium) under the ``Excellence of Science -- EOS" -- be.h project n.\ 30820817; the Beijing Municipal Science \& Technology Commission, No. Z191100007219010; the Ministry of Education, Youth and Sports (MEYS) of the Czech Republic; the Deutsche Forschungsgemeinschaft (DFG) under Germany's Excellence Strategy -- EXC 2121 ``Quantum Universe" -- 390833306; the Lend\"ulet (``Momentum") Program and the J\'anos Bolyai Research Scholarship of the Hungarian Academy of Sciences, the New National Excellence Program \'UNKP, the NKFIA research grants 123842, 123959, 124845, 124850, 125105, 128713, 128786, and 129058 (Hungary); the Council of Science and Industrial Research, India; the HOMING PLUS program of the Foundation for Polish Science, cofinanced from European Union, Regional Development Fund, the Mobility Plus program of the Ministry of Science and Higher Education, the National Science Center (Poland), contracts Harmonia 2014/14/M/ST2/00428, Opus 2014/13/B/ST2/02543, 2014/15/B/ST2/03998, and 2015/19/B/ST2/02861, Sonata-bis 2012/07/E/ST2/01406; the National Priorities Research Program by Qatar National Research Fund; the Ministry of Science and Higher Education, project no. 02.a03.21.0005 (Russia); the Programa Estatal de Fomento de la Investigaci{\'o}n Cient{\'i}fica y T{\'e}cnica de Excelencia Mar\'{\i}a de Maeztu, grant MDM-2015-0509 and the Programa Severo Ochoa del Principado de Asturias; the Thalis and Aristeia programs cofinanced by EU-ESF and the Greek NSRF; the Rachadapisek Sompot Fund for Postdoctoral Fellowship, Chulalongkorn University and the Chulalongkorn Academic into Its 2nd Century Project Advancement Project (Thailand); the Kavli Foundation; the Nvidia Corporation; the SuperMicro Corporation; the Welch Foundation, contract C-1845; and the Weston Havens Foundation (USA).
\end{acknowledgments}

\bibliography{auto_generated}

\cleardoublepage \appendix\section{The CMS Collaboration \label{app:collab}}\begin{sloppypar}\hyphenpenalty=5000\widowpenalty=500\clubpenalty=5000\vskip\cmsinstskip
\textbf{Yerevan Physics Institute, Yerevan, Armenia}\\*[0pt]
A.M.~Sirunyan$^{\textrm{\dag}}$, A.~Tumasyan
\vskip\cmsinstskip
\textbf{Institut f\"{u}r Hochenergiephysik, Wien, Austria}\\*[0pt]
W.~Adam, F.~Ambrogi, T.~Bergauer, M.~Dragicevic, J.~Er\"{o}, A.~Escalante~Del~Valle, R.~Fr\"{u}hwirth\cmsAuthorMark{1}, M.~Jeitler\cmsAuthorMark{1}, N.~Krammer, L.~Lechner, D.~Liko, T.~Madlener, I.~Mikulec, F.M.~Pitters, N.~Rad, J.~Schieck\cmsAuthorMark{1}, R.~Sch\"{o}fbeck, M.~Spanring, S.~Templ, W.~Waltenberger, C.-E.~Wulz\cmsAuthorMark{1}, M.~Zarucki
% ...
\vskip\cmsinstskip
\dag: Deceased\\
1:  Also at Vienna University of Technology, Vienna, Austria\\
% ...
\end{sloppypar}
\end{document}